\documentclass[prc,preprint,nofootinbib,superscriptaddress,showkeys,showpacs,
   preprintnumbers,amsmath,amssymb,aps,floatfix]{revtex4-1}
\usepackage{graphicx}
\usepackage{epsfig}
\usepackage{amsfonts}
\usepackage{amsmath}
\usepackage{amssymb}
\usepackage[sort&compress]{natbib}
\usepackage{dcolumn}
\usepackage{multirow}
\usepackage{bigstrut}
\usepackage{type1cm}
\usepackage{placeins}
\usepackage{rotating}
\usepackage{slashed}

\begin{document}

\preprint{\today}

\title{Concurrent tests of Lorentz invariance in $\mathbf\beta$-decay experiments}

\author{K. K. Vos}
\author{H. W. Wilschut}
\author{R. G. E. Timmermans}

\affiliation{Van Swinderen Institute for Particle Physics and Gravity, University of Groningen, Nijenborgh 4,
                   9747 AG Groningen, The Netherlands}

\date{\today}
\vspace{3em}

\begin{abstract}
Modern experiments on neutron and allowed nuclear $\beta$ decay search for new semileptonic interactions, beyond the
``left-handed'' electroweak force. We show that ongoing and planned $\beta$-decay experiments, with isotopes at rest and in
flight, can be exploited as sensitive tests of Lorentz invariance. The variety of correlations that involve the nuclear spin, the direction
of the emitted $\beta$ particle, and the recoil direction of the daughter nucleus allow for relatively simple experiments that give direct
bounds on Lorentz violation. The pertinent observables are decay-rate asymmetries and their dependence on sidereal time. We
discuss the potential of several asymmetries that together cover a large part of the parameter space for Lorentz violation in the
gauge sector. High counting statistics is required.
\end{abstract}
\maketitle

{\it Motivation.\/}---
$\beta$ decay is a recognized probe of symmetry violation in the electroweak interaction. Because of the wide choice of $\beta$
emitters and the various observables that can be measured with high precision, one can select isotopes that are tailored to
specific searches for particle physics beyond the Standard Model (SM)~\cite{Sev11,Cir13,Nav13,Vos15b}.
Over the years, strong limits were put on scalar, right-handed vector and axial vector, and tensor contributions to the semileptonic
process $d\rightarrow u+e^-+\overline{\nu}_e$. Recently, it was shown that $\beta$ decay is moreover a unique laboratory for testing
Lorentz invariance in the weak gauge~\cite{Vos15b,Noo13a,Noo13b,Vos15a} and neutrino \cite{Dia13,Dia14} sectors. Such
studies are strongly motivated by ideas how to unify the SM and general relativity in a theory of ``quantum gravity''~\cite{Lib13,Tas14}.
We demonstrate here that  ongoing and planned $\beta$-decay experiments can, with moderate modifications in the setup and
data analysis, be exploited to improve the existing limits on Lorentz violation.

We base our studies on the theoretical framework for Lorentz and CPT violation developed in Refs.~\cite{Noo13a,Noo13b} for
$\beta$ decay and in Ref.~\cite{Vos15a} for orbital electron capture. It covers effects from {\it e.g.} a modified low-energy
$W$-boson propagator $\langle W^{\mu+}W^{\nu-}\rangle=-i(g^{\mu\nu}+\chi^{\mu\nu})/M_W^2$.
%and vertex $\Gamma^\mu=(g^{\mu\nu}+\chi^{\mu\nu})\gamma_\nu$.
The tensor  components $\chi^{\mu\nu}$ were  limited with data on allowed  \cite{Wil13,Mul13,Syt15,Bod14} and forbidden
\cite{Noo13b} $\beta$ decay, pion decay \cite{Alt13b, Noo14}, nonleptonic kaon decay \cite{Vos14}, and muon decay \cite{Noo15}.
The best upper bounds %of order $\mathcal{O}(10^{-6})$-$\mathcal{O}(10^{-8})$
were derived from experiments on forbidden $\beta$ decays \cite{Noo13b}, while a first experiment on allowed $\beta$ decay
with polarized nuclei gave additional, partly complementary information \cite{Mul13,Syt15}. These results were translated into
bounds on Higgs- and $W$-boson parameters of the Standard Model Extension (SME)~\cite{Col97,Col98,Kos11}, the general
effective field theory for Lorentz and CPT violation at low energies.

The allowed-$\beta$-decay rate with Lorentz violation was derived in Ref.~\cite{Noo13a}. Compared to ordinary
$\beta$ decay, it contains additional, frame-dependent correlations between the momenta and spins of the nuclei and leptons
and the tensor $\chi$. The correlations involve linear combinations of the components $\chi^{\mu\nu}$, depending on the type
of $\beta$ decay, Fermi, Gamow-Teller, or mixed. %{\em cf.} Table I of Ref.~\cite{Noo13a}.
While many of these correlations are hard to measure, a few appear relatively straightforward. We discuss a number of
experiments on neutron and allowed nuclear $\beta$ decay that can give competing bounds on Lorentz violation. The pertinent
observables are all rather simple asymmetries recorded with sidereal-time stamps. We also consider the $\beta$ decay of nuclei
in flight, {\it e.g.} at proposed $\beta$-beam facilities, as a way to increase the sensitivity. We end with recommendations how to
further explore Lorentz violation in weak decays.

{\it Decay rate.\/}---
We assume that Lorentz violation comes from propagator corrections
%\footnote{Lorentz-violating vertex corrections also require a modification of the lepton spinors.}
and neglect momentum-dependent terms in $\chi$, which are suppressed by powers of the $W$-boson mass. Hermiticity of the
Lagrangian then implies that $\chi^{\mu\nu}=(\chi^*)^{\nu\mu}$.  We also neglect here terms with only neutrino-momentum
or neutrino-spin correlations, which are important in electron capture~\cite{Vos15a}, but in $\beta$ decay do not contain more
information than the easier to measure $\beta$-particle correlations. In addition, we ignore for the moment terms proportional
to the spin factor $\Lambda^{(2)}$ \cite{Noo13a}, which is associated with higher-order spin correlations ($\Lambda^{(2)}=0$
for unpolarized and spin-1/2 nuclei).

With these simplifications the $\beta$-decay rate \cite{Noo13a}, in the rest frame of the parent nucleus, reduces to
($\hbar=c=1$)
%\footnote{In Eq.~(18) in Ref.~\cite{Noo13a} it should read $w_3^l=-x\chi_r^ {0l}+\breve{g}(\chi_r^{0l}+\tilde{\chi}_i^l)$.
%This also corrects Eq.~(38) in Ref.~\cite{Noo13a}.}
\begin{eqnarray}
dW & = &  dW_0\; \left\{1 + 2a\chi_r^{00} +2\left(-a\chi_r^{0l} + \breve{g}\tilde{\chi}_i^{l}\right)\frac{p_e^l}{E_e}\right. \notag \\
&& +\left. \left(\left[a+2\breve{a}\chi_r^{00}\right] \delta_{lm} -4\breve{g}\chi_r^{lm}\right)\frac{p_e^lp_\nu^m}{E_eE_\nu}
       +2a\chi_i^{0k}\frac{(\vec{p}_e\times\vec{p}_\nu)^k}{E_eE_\nu} \right. \notag \\
&& +\left. \frac{\langle {J^k}\rangle}{J} \left(-2\breve{L}\tilde{\chi}_i^k
      +\left[\left(A+B\chi_r^{00}\right)\delta_{kl}-B\chi_r^{kl}\right] \frac{p_e^l}{E_e} \right)
      -A\chi_i^{0k}\frac{(\langle \vec{J}\,\rangle \times \vec{p}_e)^k}{JE_e}\right\} \ ,
\label{eq:decayrate}
\end{eqnarray}
where $dW_0=|\vec{p}_e|E_e(E_e-E_0)^2dE_e d\Omega_e  d\Omega_\nu  F(E_e,\pm Z)\xi/(2\pi)^5$, 
$\vec{p}_{e(\nu)}, E_{e(\nu)}$ are the momentum and energy of the $\beta$ particle (electron or positron) and neutrino, and
$\langle\vec{J}\,\rangle$ is the expectation value of the spin of the parent nucleus. $F(E_e,\pm Z)$ is the usual Fermi function, with
$Z$ the atomic number of the daughter nucleus, and the upper (lower) sign holds for $\beta^{-(+)}$ decays; $\xi=2C_V^2\langle 1\rangle^2
+2C_A^2\langle\sigma\rangle^2$. The subscripts $r$ and $i$ denote the real and imaginary parts of $\chi=\chi_r+i\chi_i$,
$\tilde{\chi}_i^k = \epsilon^{klm}\chi_i^{lm}$, %=2\chi_i^{lm}$,
and $k,l,m$ are spatial directions. The coefficients $a$, $A$, and $B$
are standard in $\beta$ decay \cite{Her01,Sev06}, while $\breve{a}$, $\breve{g}$, and $\breve{L}$ multiply correlations that are
Lorentz violating~\cite{Noo13a}. They are defined by
\begin{subequations}
\begin{eqnarray}
a & = & \left(1-\tfrac{1}{3}\varrho^2\right)/\left(1+\varrho^2\right) \ , \\
A & = & \left(\mp\lambda_{JJ'} \varrho^2 - 2\delta_{JJ'} \sqrt{J/(J+1)}\varrho\right)/\left(1+\varrho^2\right) \ , \\
B & = & \left(\pm \lambda_{JJ'} \varrho^2 - 2\delta_{JJ'} \sqrt{J/(J+1)}\varrho\right)/\left(1+\varrho^2\right)  \ , \\
   \breve{a} & = & \left(1+\tfrac{1}{3}\varrho^2\right)/\left(1+\varrho^2\right) \ , \\
   \breve{g} & = & \tfrac{1}{3}\varrho^2/\left(1+\varrho^2\right)  \ , \\
  \breve{L}  & = & \pm\tfrac{1}{2}\lambda_{JJ'} \varrho^2/\left(1+\varrho^2\right) \ ,
\end{eqnarray}
\end{subequations}
where $\varrho = |M_{GT}| C_A/(|M_F| C_V)$ is the ratio between the Gamow-Teller and Fermi matrix elements.
 The value of the spin factor $\lambda_{JJ'}$, where $J\,(J')$ is the initial (final) nuclear spin,
is $\lambda_{JJ'}=1$ for $J'=J-1$, $1/(J+1)$ for  $J'=J$, and $-J/(J+1)$ for $J'=J+1$.
%Neutron decay, a $J=J'=\tfrac{1}{2}$ transition, has $\lambda_{JJ'}=\tfrac{2}{3}$. 
 
{\it Observables.\/}---
\begin{table}\centering
\begin{tabular}{ccccc} \hline\hline
$X_r^{00}$ & $X_r^{0l}$ & $X_r^{kl}$ & $X_i^{0l} $ &  $\tilde{X}_i^k$% =2X_i^{lm}$
\\ \hline	
    $10^{-6}$   &     $10^{-8}$  &    $10^{-6}$   &          $-$          &  $10^{-8}$  \\
 \hline\hline
\end{tabular}
\caption{Statistical precision for the components $X^{\mu\nu}$ required to compete with the existing upper bounds
from forbidden $\beta$ decay \cite{Noo13b}. The components $X_i^{0l}$ are unconstrained at present.}
\label{tab:limits}
\end{table} 
There are 15 independent tensor components $\chi^{\mu\nu}$. %of which $\chi_i^{0l}$ are still unconstrained. 
It is standard to translate the tensor $\chi$ to the Sun-centered reference frame, in which it is denoted by $X$, and
report limits for the components $X^{\mu\nu}$ \cite{Kos11}.
%where $\chi^{\mu\nu}=R^\mu_\varrho R^\nu_\sigma X^{\varrho\sigma}$, where $R$ is the appropriate rotation \cite{Noo13a}.
The best upper bounds on (linear combinations of) $X^{\mu\nu}$ are $\mathcal{O}(10^{-6})$-$\mathcal{O}(10^{-8})$,
derived \cite{Noo13b} from pioneering forbidden-$\beta$-decay experiments \cite{New76,Ull78} that used strong sources.
In case there are accidental cancellations, the bounds on the individual components could be significantly weaker and range
from $\mathcal{O}(10^{-4})$-$\mathcal{O}(10^{-6})$ \cite{JacobPhD}. The order-of-magnitude precision required to improve
the existing bounds on the various components $X^{\mu\nu}$ is summarized in Table~\ref{tab:limits}. A statistical precision
of $10^{-n}$ requires at least $\mathcal{O}(10^{2n})$ events. This would require one year of data taking with a source of 1 Curie
for an experiment of the type performed in Ref.~\cite{New76}. An alternative option is electron capture, which allows experiments
at high rates and low dose~\cite{Vos15a}. We focus here on the possibilities to improve the existing bounds in allowed $\beta$ decay.

From Eq.~\eqref{eq:decayrate} we derive asymmetries that are proportional to specific components $\chi^{\mu\nu}$. Asymmetries 
are practical to measure and ideal to control systematic errors. Expressed in terms of $X^{\mu\nu}$, they oscillate in time
with the sidereal rotation frequency $\Omega=2\pi/(23{\rm h}\,56{\rm m})$ of Earth and depend on the colatitude $\zeta$ of the site of
the experiment. These sidereal-time variations of the observables are a unique feature of Lorentz violation, and help to separate the
desired signal from systematic errors. They also distinguish Lorentz violation from effects due to {\it e.g.}  scalar or tensor interactions,
which would produce deviations from SM predictions that are independent of Earth's orientation.

($i$) 
The simplest way to study Lorentz violation is to integrate over the neutrino direction and measure the dependence of the decay rate
on the direction of the $\beta$ particle. The highest sensitivity can be reached in pure Fermi or Gamow-Teller decays. For Fermi
decays, the experimental observable is the asymmetry
\begin{equation}
A_{F} = \frac{W_F^+ - W_F^-}{W_F^+ + W_F^-} = -2 \chi_r^{0l} \beta\hat{p}_e^l \ ,
\label{eq:fermias}
\end{equation}
where $\beta= |\vec{p}_e|/E_e$ and $W_F^{\pm}$ is the rate of $\beta$ particles measured in the $\pm\hat{p}_e$-direction.
For Gamow-Teller decays of unpolarized nuclei, the analogous asymmetry is
\begin{equation}\label{eq:gtas}
A_{GT} = \frac{W_{GT}^+-W_{GT}^-}{W_{GT}^++W_{GT}^-} = \tfrac{2}{3} \left(\chi^{0l}_r + \tilde{\chi}_i^l\right) \beta\hat{p}_e^l \ .
\end{equation}
These two asymmetries are complementary and give direct bounds on $\chi^{0l}_r$ and $\tilde{\chi}_i^l$. Mixed decays are slightly
less sensitive, {\em e.g.} for neutron $\beta$ decay, with $\varrho=\sqrt{3}\,C_A/C_V$, where $C_A/C_V\simeq -1.275$ \cite{Men13,Mun13},
the asymmetry is $A_{n}=(0.21\chi^{0l}_r+0.55\tilde{\chi}_i^l) \beta\hat{p}_e^l$.

\begin{figure}
\centering
\includegraphics[width=0.60\textwidth]{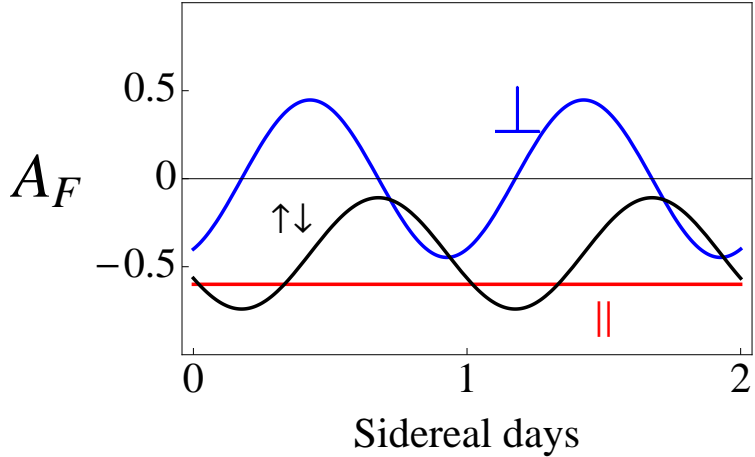}
\caption{The sidereal time dependence of the asymmetry $A_F$ in Eq.~\eqref{eq:fermias}, for $X_r^{0x}=0.1$, $X_r^{0y}=0.2$,
$X_r^{0z}=0.3$, and colatitude $\zeta=45^\circ$. For $\beta$ particles observed parallel $(\parallel)$ to Earth's rotation axis $A_F$
is constant. Observation in the $\uparrow\downarrow$ (up-down) direction or perpendicular $(\perp)$ to the rotation axis results in an
oscillation of $A_F$ with sidereal time.}
\label{fig:asymm}
\end{figure}

Figure~\ref{fig:asymm} illustrates the sidereal-time dependence of the asymmetry $A_F$ for three different observation directions.
When the $\beta$ particles are detected parallel to Earth's rotation axis, no oscillation is observed. Observation of the $\beta$ particles
perpendicular to the rotation axis, {\it i.e.} east-west, gives a sidereal-time variation. When the $\beta$ particles are observed in the
up-down ($\uparrow\downarrow$) direction, this oscillation has a constant offset.  Systematic errors can result in a finite offset, and
therefore observation in the direction perpendicular to the rotation axis is favored. The asymmetries should preferably be measured
in a rotating setup~\cite{New76} to reduce systematic errors. Alternatively, a multi-detector setup with appropriate symmetry can
exploit the full polar and azimuthal dependence as shown in Fig.~\ref{fig:asymm}, while reducing the counting rates of the individual
detectors. An experiment with a duration of one year can use diurnal variations to reduce systematic errors.

There are ongoing efforts to improve the bounds on tensor currents in $\beta$ decay. A promising observable for this purpose is the
energy spectrum of the $\beta$ particles~\cite{Sev14}. The Gamow-Teller decays of $^6$He \cite{Kne11,Nav14} and $^{45}$Ca \cite{Sev14}
are under consideration. Such experiments require high statistics and accuracy. The $^6$He facility promises to produce
$10^{10}$ particles/s, but it remains to be seen how such a beam can be used for Lorentz-violation measurements~\cite{Nav14}.
Isotopes such as $^{32,33}$P, $^{35}$S, and $^{63}$Ni are also of interest, because they have clean ground-state-to-ground-state
$\beta^-$-transitions and low $Q$-values. For example, conveniently-shaped $^{63}$Ni sources of 1GBq are commercially available.
Such sources have minimal contributions of secondary radiation that can complicate the measurements.  Moreover, strong sources
can be produced in reactors. For the Fermi asymmetry $A_F$, any of the superallowed $0^+ \rightarrow 0^+$ decays
\cite{Har09,Har15} can be considered. We recommend that in these experiments the asymmetries $A_F$ of Eq.~\eqref{eq:fermias} and
$A_{GT}$ of Eq.~\eqref{eq:gtas} are measured concurrently, with sidereal-time stamps.
%Neutron $\beta$ decay is another clear option. High-intensity neutron sources are available at various facilities. 

($ii$)
With polarized nuclei one can measure the correlations that involve the nuclear spin. The simplest of these is the spin asymmetry 
\begin{equation}\label{eq:asspin}
A_{J} = \frac{W^\uparrow -   W^\downarrow}{W^\uparrow +   W^\downarrow} = -2\breve{L} \tilde{\chi}_i^k P\hat{J}^k \ ,
\end{equation}
where $\hat{J}$ is the unit vector in the direction of the parent spin, $P$ is the degree of nuclear polarization, and
$W^{\uparrow (\downarrow)}$ is the integrated decay rate in the $\pm\hat{J}$-directions. For pure Gamow-Teller decays,
$\breve{L} = \tfrac{1}{2}B =-\tfrac{1}{2}A$.  Isotopes for which $\lambda_{JJ'}=1$ are optimal.

The first dedicated experiment to search for Lorentz violation in allowed $\beta$ decay measured $A_J$ in the $\beta^+$ decay
of $^{20}$Na \cite{Mul13}. The result of the most recent measurement is $\left|\tilde{\chi}_i^{x,y}\right|< 5\times10^{-4}$ with
90$\%$ confidence~\cite{Syt15}. Data for polarized-neutron decay are currently being analyzed \cite{Bod14,Koz11}.
When the sidereal-time dependence of $A_J$ is measured, it is not necessary to know $A$ and $P$ with high precision.
If the polarization is not exactly equal in the two directions, $A_J$ will show an offset, which is independent of the sidereal
frequency as long as the polarization can be kept independent of $\Omega$. Still, a measurement of the $\beta$ asymmetry
$A_{GT}$, as discussed above, is probably preferable for improving the bounds on $\tilde{\chi}_i^k$. 

($iii$)
The components $\chi_i^{0k}$, for which there are no bounds available yet, can be accessed through the correlations of
$\hat{J}\times\hat{p}_e$ or $\hat{p}_e\times\hat{p}_\nu$ and a component of $\chi$. The first correlation can be measured with
the asymmetry  
\begin{equation}\label{eq:abetanu}
A_{\beta\nu} = \frac{W_{L}^\uparrow W_{R}^\downarrow -  W_{R}^\uparrow W_{L}^\downarrow}{W_{L}^\uparrow W_{R}^\downarrow +  
W_{R}^\uparrow W_{L}^\downarrow} = 4\left(a\chi_i^{0k}\epsilon^{klm} -2\breve{g}\chi_r^{lm}\right) \beta\hat{p}_e^l\hat{p}_\nu^m \ ,
\end{equation}
where $W_{L,R}$ is obtained by measuring the $\beta$ particles in the opposite left ($L$) and right ($R$) $\hat{p}_e$-directions, while
the recoiling nucleus is detected in the perpendicular $\uparrow(\downarrow)$ direction. For this asymmetry, pure Fermi decays, with
$a=1$ and $\breve{g}=0$, are preferred. Experiments that measure both the $\beta$ and the neutrino direction are thus of interest.
Ref.~\cite{Gor05} {\em e.g.} reports a search for a deviation from the SM prediction $a=1$ for the $\beta$-$\nu$ correlation in $^{38m}$K,
with an error on $a$ of order ${\cal O}(10^{-3})$, which would be the corresponding limit for $\chi_i^{0k}$.

With polarized nuclei, $\chi_i^{0k}$ can be measured from the asymmetry between the nuclear spin and the $\beta$ particle,
\begin{equation}\label{eq:aJbeta}
A_{J\beta} = \frac{W_{L}^\uparrow W_{R}^\downarrow -   W_{R}^\uparrow W_{L}^\downarrow}{W_{L}^\uparrow W_{R}^\downarrow + 
W_{R}^\uparrow W_{L}^\downarrow} = -2 \left(A \chi_i^{0m} \epsilon^{mkl} + B\chi_r^{kl}\right) P\hat{J}^k \beta\hat{p}_e^l  \ ,
\end{equation}
where now $W_{L,R}$ is the rate with the $\beta$ particles in the opposite left ($L$) and right ($R$) $\hat{p}_e$-directions and the
nuclei polarized in the perpendicular $\uparrow(\downarrow)$ $\hat{J}$-direction. Equation~(\ref{eq:aJbeta}) holds for Gamow-Teller
and mixed decays. Gamow-Teller decays with $\lambda_{JJ'}=1$ are preferred.

The bounds from forbidden $\beta$ decay give $\left|\chi_r^{kl}\right|<\mathcal{O}(10^{-6})$ \cite{Noo13b}. A measurement of
$A_{J\beta}$ or $A_{\beta\nu}$ with a precision lower than $10^{-6}$, therefore, translates to a bound on $\chi^{0k}_i$.
The sidereal-time variation of $A_{J\beta}$ and $A_{\beta\nu}$ is similar to that shown in Fig.~\ref{fig:asymm}. To reduce
systematic errors $\hat{J}\times\hat{p}_e$ or $\hat{p}_e\times\hat{p}_\nu$ should point perpendicular to Earth's rotation axis. 
$A_{J\beta}$ can possibly be obtained in polarized-neutron decay by reanalyzing the data of Ref.~\cite{Bod14}. Measuring
the asymmetries better than $10^{-6}$ requires coincident event rates exceeding $3\times 10^4/$s for a year, but will
then also improve the bounds on $\chi_r^{kl}$.

{\it Exploiting Lorentz boosts.\/}---
So far we discussed $\beta$ decay of nuclei at rest. The required event rate in these measurements is a challenge.
In forbidden $\beta$ decays one can benefit from an enhancement of Lorentz violation of one order of magnitude~\cite{Noo13b}.
A much larger enhancement can be obtained when the decaying particle is in flight.  Consider specifically the total decay rate,
which in the rest frame depends only on the isotropic term in Eq.~\eqref{eq:decayrate},
\begin{equation}\label{eq:chir00}
   W/W_0 = 1 + 2a \chi_r^{00} \ ,
\end{equation}
where $W_0$ is the SM decay rate and $a=1$ $(-1/3)$ for Fermi (Gamow-Teller) decays. The component $\chi_r^{00}$
can {\em e.g.} be measured from the ratio between the longitudinal $\beta$ polarization,
$P_\beta = \left(1+2a\chi_r^{00}\right)G\beta$, for Fermi and Gamow-Teller decays, where $G=\mp 1$ \cite{Her01,Sev06}.
Comparing the best value $P_F/P_{GT}=1.0010(27)$ \cite{Wic86,Car90} to the SM prediction $P_F/P_{GT} =1$
gives $-1.3\cdot 10^{-3}<X^{00}_r<2.0\cdot 10^{-3}$ with 90\% confidence, which is a much weaker bound than the
one obtained for forbidden $\beta$ decay~\cite{Noo13b}, and hard to improve with nuclei at rest.
%$\chi_r^{00}$ can be measured by comparing decay rates at different $\gamma$, as was done for muon decay~\cite{Noo15}.

The decay rate in flight depends on the velocity $\vec{v}=v\hat{v}$ of the nucleus that results from a Lorentz boost.
In terms of the components $X^{\mu\nu}$ in the Sun-centered frame one has
%\begin{equation}
$\chi_r^{00} = \gamma^2 \left(X_r^{00}-2v X_r^{0l}\hat{v}_l+X_r^{kl}v^2\hat{v}_k\hat{v}_l\right)$,  
%\end{equation}
where $\gamma=1/\sqrt{1-v^2}$ is the Lorentz factor. When the velocity $\hat{v}$ is perpendicular to Earth's rotation
axis (east-west) one finds
\begin{eqnarray} 
\chi_r^{00} & = & \gamma^2 \left(X_r^{00}+\tfrac{1}{2}v^2\left[X_r^{xx}+X_r^{yy}\right]+2v X_r^{0x}\sin\Omega t
                             -2v X^{0y}_r \cos\Omega t  \right. \nonumber \\
                     && \left. - v^2 X_r^{xy} \sin2\Omega t - \tfrac{1}{2}v^2\left[X_r^{xx}-X_r^{yy}\right] \cos2\Omega t \right) \ ,
\label{eq:Xrkl}
\end{eqnarray}
which is enhanced by a factor $\gamma^2$.
The components $X_r^{\mu\nu}$ can be fitted to the sidereal-time dependence of the measured decay rate. Alternatively,
one can measure the decay rate at time $t$ and 12 hours later, and isolate $X_r^{0l}$ via the ``asymmetry''
\begin{equation} \label{eq:At}
A_{t} = \frac{W(\Omega t)-W(\Omega t + \pi)}{W(\Omega t)+ W(\Omega t+\pi)}
         = 4a\,v\gamma^2 \left(X^{0x}_r \sin\Omega t -X^{0y}_r \cos\Omega t\right) \ ,
\end{equation}
while $X_r^{kl}$ can be obtained by measuring at intervals of 6 hours, with
\begin{eqnarray}\label{eq:A2t}
A_{2t} & = & \frac{W(\Omega t)-W(\Omega t + \tfrac{1}{2}\pi)+W(\Omega t+ \pi)-
                      W(\Omega t + \frac{3}{2}\pi)}{W(\Omega t)+W(\Omega t + \tfrac{1}{2}\pi)+
                      W(\Omega t+ \pi)+W(\Omega t + \frac{3}{2}\pi)} \nonumber \\
& = & - a\,v^2 \gamma^2 \left(\left[X_r^{xx}-X_r^{yy}\right] \cos 2\Omega t + 2 X^{xy}_r \sin 2\Omega t\right) \ ,
\end{eqnarray}
which oscillates only with the double frequency $2\Omega$.

The $\gamma^2$ enhancement in Eqs.~(\ref{eq:Xrkl}), (\ref{eq:At}), and (\ref{eq:A2t}) can be exploited at a $\beta$-beam facility
planned for neutrino physics \cite{Lin10}. A good nucleus for such a facility is $^6$He, for which the production rates are estimated
at $10^{12}/$s with $\gamma=100$ \cite{Aut03}. A possible setup for a $\beta$-beam facility that uses the proton synchrotrons
at CERN is discussed in Refs.~\cite{Aut03, Lin10}.

Of course, any weakly-decaying particle in flight can be used, provided the coefficient $a$ in Eq.~(\ref{eq:chir00}) can
be calculated reliably. Nonleptonic decays of strange hadrons such as kaons are problematic \cite{Vos14}, but decays
of heavy quarks do not have this drawback. Leptonic and semileptonic decays are clearly preferable.  For fast-moving
pions \cite{Alt13a} bounds of $\mathcal{O}(10^{-4})$ on $\chi^{\mu\nu}$ were obtained~\cite{Alt13b}. Semileptonic kaon
decays have been studied at the SPS at CERN \cite{Bat07} with $\gamma\simeq 150$ and will be part of the background
in the NA62 experiment. LHCb, designed to observe decays at $\gamma\gtrsim10$, is serendipitously oriented
perpendicular to Earth's rotation axis. For all accelerator studies, the precise normalization of the decay rate as
function of sidereal time  is necessary for a concurrent test of Lorentz invariance.

{\it $\beta$-$\gamma$ correlations.\/}---
We have only considered cases where the anisotropic decay rate is observed in the emission direction of the $\beta$ particles
and/or is associated with the polarization direction of the parent nucleus. The anisotropy can also be observed from $\gamma$
rays when an excited state in the daughter nucleus is populated. In Gamow-Teller transitions the daughter nucleus is left in a polarized
state that reflects the degree of anisotropy of the emission. When measuring the $\gamma$-decay angular distribution this
anisotropy can be observed as a residual alignment. Inspection of Eq.~(\ref{eq:decayrate}) shows that this will be the case
for the term $-2\breve{L} \tilde{\chi}_i^k\hat{J}^k$.  Clearly, such a measurement will have lower sensitivity compared with
the direct measurements discussed above. The last line of Eq.~(\ref{eq:decayrate}) can also be accessed by measuring
$\beta$-$\gamma$ correlations. The last term is relevant because it contains the ``missing'' components $\chi_i^{0k}$.
In this case the lower sensitivity may be compensated by an efficient setup. To obtain the actual expressions and the corresponding asymmetries the terms proportional to $\Lambda^{(2)}$ \cite{Noo13a} have to be added to
Eq.~(\ref{eq:decayrate}). The evaluation depends on the particular details of detection method and will be considered when the
need arises.

{\it Conclusion.\/}---
The breaking of Lorentz invariance in the weak interaction can be probed in relatively simple allowed-$\beta$-decay
experiments. We propose to measure a number of decay-rate asymmetries as function of sidereal time, which together can
constrain all Lorentz-violating gauge components. Measurements of the $\beta$-decay asymmetry in Fermi and Gamow-Teller
decays, Eq.~\eqref{eq:fermias} and Eq.~\eqref{eq:gtas}, give direct bounds on $\chi_r^{0l}$ and $\tilde{\chi}_i^k$. The most
complicated experiments require the measurement of a correlation between two observables, as in Eq.~\eqref{eq:abetanu}
or Eq.~\eqref{eq:aJbeta}. The components $\chi_i^{0k}$ are still unconstrained and these measurements will give the first
bounds. In addition, we point out the potential of $\beta$ beams and LHCb for tests of Lorentz invariance. Ultimately, the experiments
should aim to improve the existing forbidden-$\beta$-decay limits starting at $\mathcal{O}(10^{-6})$, which requires high-intensity
sources and excellent control of systematic uncertainties. As we have shown, this can go hand-in-hand with high-precision
allowed-$\beta$-decay experiments that search for new semileptonic physics. Such efforts are, therefore, of considerable
general interest.

{\it Acknowledgments.\/}---
This investigation grew out of discussions at the 33$^{\rm rd}$ Solvay Workshop on Beta Decay Weak Interaction Studies
in the Era of the LHC (Brussels, September 3-5, 2014). We thank J. Noordmans for helpful discussions. The research was
supported by the Dutch Stichting voor Fundamenteel Onderzoek der Materie (FOM).

\end{document}